\documentclass[pra,amsfonts,preprint,showkeys]{revtex4}

\begin{document}

\title{Quantum randomness and value indefiniteness}

\author{Cristian S. Calude\footnote{Work done at the University of Technology Vienna: the support of the Institute for Theoretical Physics is gratefully acknowledged.}}
\email{cristian@cs.auckland.ac.nz}
\homepage{http://www.cs.auckland.ac.nz/~cristian/}
\affiliation{Department of Computer Science,
The University of Auckland,
Private Bag 92019,
Auckland,
New Zealand}
\author{Karl Svozil}
\email{svozil@tuwien.ac.at}
\homepage{http://tph.tuwien.ac.at/~svozil}
\affiliation{Institute for Theoretical Physics, University of Technology Vienna,
Wiedner Hauptstra\ss e 8-10/136, A-1040 Vienna, Austria}

\begin{abstract}
As computability implies value definiteness, certain sequences of quantum outcomes cannot be computable.
\end{abstract}


\maketitle
\thispagestyle{empty}
\section{Conceptualisation}

It certainly would be fascinating to pinpoint the time
of the emergence of the notion that certain quantum processes, such as the decay
of an excited quantum state, occurs principally and irreducibly at random;
and how long it took
to become the dominant way of thinking about them after almost two centuries
of quasi-rationalistic dominance.
Bohr's and Heisenberg's influence has been highly recognised and has prevailed,
even against the strong rationalistic and philosophic objections
raised by, for instance, by Einstein and Schr\"odinger \cite{jammer:66,jammer1}.
Of course, one of the strongest reasons for this growing acceptance
of quantum randomness has been the factual inability to go ``beyond'' the
quantum in any manner which would encourage new phenomenology and might
result in any hope for a progressive quasi-classical research program  \cite{lakatosch}.

Here we intend to discuss quantum randomness and its connection with
quantum value indefiniteness.
Bell \cite{bell-66,bell,bell-87,pitowsky-89a}, Kochen and Specker (KS) \cite{kochen1},
as well as Greenberger, Horne and Zeilinger (GHZ) \cite{ghz,ghsz,panbdwz}
contributed to the evidence that the mere concept of
coexistence of certain elements of physical reality  \cite{epr}
results in a complete contradiction.
In this view, speculations about the ``reasons'' for certain outcomes of experiments
are necessarily doomed; just because of the simple fact that
any such rational reason is provably (by contradiction) impossible.

An attempt is made here to clearly spell out the issues and problems involved in
considering randomness, both with regard to the occurrence of single events,
as well as their combination into time series.
We wish to state from the beginning that we attempt to have no bias or
preference for or against randomness. While to us  it seems obvious that
any claim of non-randomness has to be confronted with the factual inability
to produce any satisfactory theory that goes beyond the quantum,
especially in view of the known no-go theorems by Bell, KS and GHZ  and others
referred to above,
it is also advisable to  keep all options open and carefully study
the types of randomness involved, and their possible ``origins,'' if any.

Usually,  the random outcome of certain quantum physical events
seems to be axiomatically postulated
from the onset; an assumption which
can be also based on elementary principles \cite{zeil-99,zeil-bruk-02}.
Here we argue that actually we can go further and
infer some properties of quantum randomness --- including the absence of effective global correlations ---
from the impossibility of value definiteness of certain quantum mechanical observables.

\subsection{Difficulties}

Consider, as two extreme cases,
the binary expansion $\pi_1\pi_2\pi_3\ldots \pi_i \pi_{i+1}  \ldots$ of pi, an ideal circle's ratio of the circumference to its diameter,
starting from, say, the 571113th billion prime number place onwards,
and compare it to a sequence generated by quantum coin tosses
$x_1x_2x_3\ldots x_i x_{i+1}  \ldots$
\cite{svozil-qct,zeilinger:qct,Cris04}.
How could anyone possibly see a difference with respect to their (non-)stochasticity?
For all practical purposes, the sequences will appear structurally identically
from a stochastic point of view,
and heuristically random. For example, both are unknown to be Borel normal; i.e.,
all finite sub-sequences $y_1y_2y_3\ldots y_N$
might be contained in them with the expected frequencies.
Indeed, it is not unreasonable to speculate that the pi sequence
might be immune to all statistical and algorithmic
tests of randomness but one:
a test against the assumption that it is the binary expansion of pi, starting from the 571113th billion prime number place onwards.

Another obstacle for the physical conceptualisation of quantum randomness
and its operationalisation in terms of physical entities
originates in the formalism upon which such
endeavours have to be based.
The formal incompleteness and independence
discovered by G\"odel, Tarski, Turing, Chaitin and others essentially renders
algorithmic proofs of randomness hopeless. We shall discuss these issues below,
but we just note that, as an example, verification of
any ``law'' describable by $k$ symbols requires times exceeding any computable
 function  of $k$ [such as the Ackermann function $A(k)$]
and could in general take also that long to be falsified.
Thus, the proof of any absence of lawful behaviour seems  provable impossible.

Randomness is an asymptotic property, that is,  it is unaffected by finite variations. This makes testing randomness extremely difficult: one has to find finite tests capable of distinguishing an infinite behaviour.


\subsection{Scenarios}

 Quantum randomness appears to occur in two different scenarios:
(i) the complete impossibility to predict or explain the occurrence of certain
{\em single} events and measurement outcomes from any kind
of operational causal connection. The hidden ``parameter models'' for the quantum phenomena which
have been proposed so far do not provide more insight for the predictions
of intrinsic observers embedded in the system;
and
(ii) the concatenation of such single quantum random events forms sequences of random bits
which can be expected to be equivalent stochastically to  white noise.
White noise carries the least correlations, as the occurrence of a particular bit value
in a binary expansion does not depend on previous or future bits of that expansion
\cite{gard-78}.

These different ways to encounter randomness ---
single random events and a concatenation thereof ---
should be perceived very differently:
in the single event case, the outcome occurs in the highly complex
environment of the quantum and its measurement apparatus, which is thereby
``folded'' into a single bit.
Repetition of the experiment does not increase
the complexity of the combined system of the quantum--measurement apparatus,
whose
repetitive
properties and behaviours are ``unfolded'' in repeated experiments.
Hence, possible biases against statistical tests may be revealed easier by considering sequences of single random outcomes.
In this note we shall thus concentrate on this second.

\subsection{Axioms for quantum randomness and degrees of randomness}
In what follows, we will assume the standard two ``axioms'' for
quantum randomness~\cite{jauch}:

\begin{itemize}
\item The single outcome
from which quantum random sequences are formed, occurs unbiased; i.e.,
for the $i$th outcome, there is a 50:50 probability for either 0 or 1:
\begin{equation}
\begin{array}{l}
\textrm{Prob}(x_i=0)
=
\textrm{Prob}(x_i=1)
={1\over 2}\raisebox{0.5ex}{.}
\end{array}
\end{equation}
\item   There is a total independence of previous history,
such that no correlation exists between $x_i$ and previous or future outcomes.
This means that the system carries no memories of previous or expectations of future events.
All outcomes are temporally ``isolated'' and
free from control, influence and determination.
They are both unbiased and self-contained.
\end{itemize}

Assume that we have a quantum experiment (using light, for example: a
photon generated by a source   beamed to a semitransparent mirror
is
{\it ideally}
reflected or transmitted with 50 per cent chance) which at each stage produces a quantum random bit, and we assume that this experiment is run for ever generating an infinite binary sequence:
\begin{equation}
\label{qrand}
X =x_1x_2x_3\cdots x_i \cdots
\end{equation}

In this scenario, the first axiom shows that the limiting frequency
of 0 and 1 in the sequence $X$ is $1/2$. Locally, we might record
significant deviations, i.e., $X$ may well start with a thousand of 1's, but in the limit these discrepancies disappear.

The ``lack of correlations''  postulated above is more difficult to understand and may easily lead to misunderstandings, hence errors.
First, {\it finite correlations} will always exist, because of the asymptotic nature of ``randomness''.
Secondly, even {\it infinite correlations} cannot be eliminated because they have been proven to exist in {\it every infinite sequence};  for example Ramsey-type correlations, see \cite{calude:02}. So, what type of correlations should be prohibited?
There are many possible choices, but the ones which come naturally to mind are ``effectively computable defined correlations.'' In other terms, correlations --- finite or infinite --- which can be detected in an effective/algorithmic way, should be excluded.

Once the nature of the two axioms of randomness has been clarified, we
can ask ourselves whether we need both axioms, that is, whether the axioms are {\it independent}. The answer is {\it affirmative} and here
is the proof.
An example of a binary sequence which satisfies
the first axiom, but not the second axiom  is Champernowne's  sequence
$0100011011000001010011 \cdots 1110000 \cdots$, which is just the  concatenation of all binary strings in quasi-lexicographical order. In this sequence 0 and 1 have limiting frequency $1/2$ (even, more, each
string of length $n$ has limiting frequency exactly $1/2^{n}$), but, of course, this sequence is computable, so  it contains infinitely many finite and infinite correlations.

It is possible to transform a sequence $Z = z_{1}z_{2}\cdots$ with no correlations and limiting frequency of 0's (and 1's) exactly $1/2$ into
a sequence which has no infinite correlations, but
the limiting frequency of 0's is $2/3$ and the limiting frequency of 1's is $1/3$:
 replace in $Z$ every 0 by 001, and every 1 by 100. This new sequence will have ``weak local correlations'' --- for example 0010 has to be followed by 01 --- but those correlations  are not global.

We stress the fact that we are interested in  ``theoretical'' sequences (\ref{qrand}) produced by an ideal quantum experiment generating randomness, not the specific results of a particular quantum device
like {\it Quantis}, \cite{Quantis}. Real devices are
prone to real-world imperfections, even watered-down by various
unbiasing methods, see \cite{Cris04}; however, our results apply in the limit to sequences generated by devices like  {\it Quantis} (see
\cite{Cris04,calude-dinneen05}).

What is the degree of ``randomness'' of the resulting white noise sequence? Theoretically there are a few possibilities, ranging from
``total randomness'' expressed mathematically by saying that the sequence is algorithmically incompressible or algorithmically random,\cite{calude:02}  to weaker and weaker possibilities: Turing-uncomputable of various degrees, but not algorithmically incompressible, Turing-computable, easy Turing-computable. Which of these possibilities actually does occur?

\section{Main results}

\subsection{Quantum value indefiniteness}

In classical physics, omniscience
manifests itself in the implicit assumption that
it is possible to know all physical properties, or to put it in the
context of the Einstein, Podolsky and Rosen argument \cite{epr},
all ``elements of physical reality'' are definite.
Classical realism assumes that these definite physical properties exist
without being experienced by any finite mind \cite{stace},
that it would not matter whether or not a particular physical observable
is measured or not; and that the outcome of any such measurement is independent of
whatever is measured alongside with it; that is, of its context.
To state it pointedly: all classical physical observables exist
simultaneously and independent of observation.

Complementarity expresses the impossibility to measure two observables,
such as the spin states of two spin-${1\over 2}$ particles
along orthogonal directions, with arbitrary precision.
But, as equivalent \cite{svozil-2001-eua} generalised urn \cite{wright}
or automaton models \cite{svozil-ql} demonstrate,
complementarity does not necessarily imply value indefiniteness.
There still could exist enough two-valued states on the associated propositional
structures to allow a faithful embedding into a Boolean algebra associated with classical physical systems.
Formally, value indefiniteness manifests itself in the ``scarcity'' or non-existence of
two-valued states --interpretable as classical truth assignments -- on all
or even merely a finite set of physical observables.
This is known as the Kochen-Specker theorem
\cite{kochen1} (for related results, see Refs.~\cite{ZirlSchl-65,Alda,Alda2,kamber64,kamber65}).
Very similar conclusions can be drawn from the impossibility to enumerate tables of results
associated with Bell-type experiments in a consistent way: no such tables could possibly reproduce
the non-classical quantum correlations \cite{peres222,krenn,svozil-krenn}.

Confronted with the impossibility to consistently assign globally defined
observables, one may assume, in an attempt to maintain realism, that the outcome of a particular experiment
depends on the other observables which are co-measured simultaneously (Bell~\cite{bell-66}, Sec.~5).
This assumption is called ``context dependence.''

Alternatively, one may depart from classical omniscience and assume that
an elementary two-state system can carry at least a single bit,
and nothing more.
The context enters in the form of the maximal operator, such that all other co-measurable operators
are functions thereof.
If a particle can be prepared only to be in a single context,
then the question quite naturally arises why the measurement
of a different context not matching the preparation context
yields any outcome at all.
Pointedly stated, it is amazing that {\it for non-matching contexts there is an outcome rather than none}.
We note that only under these circumstances, quantum randomness manifests itself,
because if the preparation and the measurement contexts match,
the measurement just renders the definite outcome associated with
the state in which the particle was prepared.
In this non-contextual view, quantum value indefiniteness
expresses the fact that no deterministic, (pre-) defined non-contextual element of physical reality
could consistently exist for observables in contexts not matching the preparation context.
This is true also if we assume some form of ``context translation'' which may introduce stochasticity
through some mechanism of interaction with the measurement device.

\subsection{From value indefiniteness to Turing-uncomputability}

Thus, we conclude, {\it no non-contextual, deterministic computation could exist which
yields such a measurement outcome}.
If one insists on some form of agent producing the outcome,
then this agent  must perform like an erratic gambler
rather than a faithful executor of a deterministic algorithm.

Restated differently, suppose
a quantum sequence hitherto considered would be computable.
In this case, the computations involved would produce a definite number associated with
a definite outcome, which in turn could be associated
with a definite element of physical reality.
Yet we know that for Hilbert spaces of dimension greater than two,
the assumption of value definiteness of all possible observables
results in a complete contradiction.
Hence, one is forced to conclude that  {\it  the assumption of computability has to be given up,
and hence the  sequence $X$ in (\ref{qrand})  is Turing-uncomputable.}

Because the class of computable sequences is countable, with probability one (even, constructively, with probability one, see \cite{calude:02}) every sequence is Turing-uncomputable. Our result stated above is much stronger: {\it no  sequence $X$  in (\ref{qrand})  is Turing-computable}. In particular, it says that any sequence $X$ cannot contain only 0's, it  cannot represent in binary the digits of the binary expansion of pi or the Champernowne sequence. More,  no  sequence $X$  can coincide with a pseudo-random sequence
(i.e., sequence obtain via  Turing machine program), a   fact alluded  to almost 50 years ago by  John von Neumann: ``Anyone who considers arithmetical methods of producing random digits is, of course, in a state of sin''.

\subsection{White noise and algorithmic incompressibility}

Uncomputability is a strong property, but it does not necessarily imply  algorithmically incompressibility. Is a sequence $X$
more similar to the typical Turing-uncomputable sequence  given by
the classification of Turing programs in halting or non-halting, \begin{equation}
\label{hseq}
H =h_1h_2h_3\cdots h_i \cdots,
\end{equation}
or to the sequence of bits of a Chaitin Omega number, the halting probability:
\begin{equation}
\label{omega}
\Omega =\omega_1 \omega_2 \omega_3\cdots \omega_i \cdots ?
\end{equation}

The sequence $H$ is defined by assigning to $h_{i}$ the value 1 if the $i$th Turing program (in some
systematic enumeration) halts, and the value 0 in the opposite case.

The sequence $\Omega$ is obtained by working with self-delimiting Turing machines (i.e. machines with prefix-free domains) by the formula:

\[\Omega = \sum_{p {\footnotesize \mbox{   halts}}} 2^{-|p|},\]
where $|p|$ denotes the length (in bits) of the program $p$ (see more in \cite{calude:02}).

Both $H$ and $\Omega$ are Turing-uncomputable. The sequence $H$ is  Turing-uncomputable, but it is also not
algorithmic incompressible. A reason is the fact that we can
effectively  compute   infinitely many exact values of $H$ by  explicitly
constructing infinitely many halting (or, non-halting) programs.
The sequence $\Omega$ is algorithmic incompressible. Both  $H$ and $\Omega$ can solve the famous Halting Problem: we need the first $2^{n}$ bits of  $H$ to solve the Halting Problem for programs $p$ of length
$|p| \le n$, but we need no more than the first  $n$ bits of $\Omega$ to solve the same problem.
The prefixes of $\Omega$ encode the same amount of information as the prefixes of $H$, but in an exponentially more compressed way.

It is not difficult to see that the argument presented below  to show that $X$
is Turing-uncomputable can be adapted to prove that {\it every infinite sub-sequence of $X$ is Turing-uncomputable.}
More formally, {\it there is no partially computable function $\varphi$
defined on an infinite set of positive integers such that if $\varphi (n)$ is defined, then $\varphi (n) = x_{n}$}. This property is called {\it bi-immunity} in the theory of computability, see Odiffredi \cite{odi:99}.

This property is shared by $\Omega$, but not by $H$.

\subsection{Some consequences}

We discuss some simple consequences of the above result.

First, no Turing machine can enumerate/compute any sub-sequence
of  $X$. This means that every given Turing machine can compute only finitely many exact bits of $X$ in the same way that every given Turing machine can compute only finitely many exact bits of $\Omega$ (in contrast with $H$). Similarly, any formal system (ZFC, for example) will be able to ``prove'' only finitely many exact values of the sequence $X$.

Secondly, the sequence $X$ is {\it not predictable}. The most clear intuition people  have about randomness is unpredictability:
the  bits of a ``random'' sequence  should be such that one cannot predict the next bit even if one knows all preceding bits.  The simplest way to model this phenomenon (see  other models in \cite{DH}) is to consider
predictions of the $(n+1)$th element of the sequence when one knows the first $n$ elements. The corresponding model is to accept as  {\it predictor} a partial computable function $Pred$ defined on  a subset of the prefixes of $X$
 with 0-1 values.  If $Pred(w)= z$ and $z= x_{|w|+1}$ we say that the bit $z$
 was predicted from  $w$.
Does there
exist a predictor $Pred$ predicting infinitely many bits of $X$? The answer is clearly negative: from $Pred$ we can construct a partially computable function $\varphi$ capable of enumerating infinitely many
values of $X$ just by enumerating the domain of $Pred$ and each
time we get $Pred(w)= z$ and $z= x_{|w|+1}$, then we put $\varphi (|w|)=z$. This leads to a contradiction.

Thirdly,  a more general result can be proved: {\it there are no
effective global (infinite) correlations between the bits of $X$.} One way to formalise this idea is to consider all possible properties between
the prefixes of $X$ that can be determined in an effective way. We can prove
the following result: {\it Every infinite relation of the form
$G =\{(u,v)\mid uv \mbox{  is a prefix of } X\}$ is not computably enumerable.}   Indeed, from $G$ we can construct the partial function
$\varphi$ as follows: to the pair $(u,v)\in G, v = v_{1}v_{2}\cdots v_{m}$ we associate the following values of $\varphi$: $\varphi (|u| + i-1)= v_{i}, i= 1, \ldots ,m. $ The function $\varphi$ is correctly defined because of the condition specified in the definition of $G$; it shows that
one can effectively enumerate  infinitely many bits of $X$, a contradiction.

\section{Summary and discussion}

We have argued that, because of the value indefiniteness encountered
in quantum mechanics, there cannot exist deterministic computations
``yielding''
infinitely many
individual quantum random bits.
We have further exploited value indefiniteness formally by stating the consequences
in terms of Turing-uncomputability for sequences of such quantum random bits.
No effectively computable global correlations can exist between the bits of a
quantum random sequence.

We have also examined,
in a theoretical manner,
the role of the second axiom of quantum randomness.
The first axiom, stochasticity,
seems more difficult to be studied from a purely theoretical point of view
--- of course, it will be extremely interesting to have results in this direction ---
but can be experimentally approached
(for example, with the help of statistically significant samples produced by {\it Quantis}).

Finally, we note that the result presented in this note says nothing about the possibility of extracting  quantum bits from the
quantum source of randomness, which, one might hope, could enhance
the power of ``real'' computation. Some impossibility results in this direction were proved in \cite{dodis-renner}.

\section*{Acknowledgement} We thank Ludwig Staiger for illuminating discussions on bi-immunity and stochasticity.



\end{document}